# Resonance frequency measurement with accuracy and stability at the $10^{-12}$ level in a copper microwave cavity below 26 K by experimental optimization

Haiyang Zhang[a,b,&], Bo Gao[a,b,&,*], Wenjing Liu[a,b,e], Changzhao Pan[c,a], Dongxu Han[d,a], Ercang Luo[a,b,e], Laurent Pitre[c,a]

[a]TIPC-LNE Joint Laboratory on Cryogenic Metrology Science and Technology, Technical Institute of Physics and Chemistry, Chinese Academy of Science, Beijing 100190, China

[b]Key Laboratory of Cryogenics, Technical Institute of Physics and Chemistry (TIPC), Chinese Academy of Sciences (CAS), Beijing 100190, China

[c]Laboratoire national de métrologie et d'essais-Conservatoire national des arts et métiers (LNE-Cnam), La Plaine-Saint Denis F93210, France

[d]School of Mechanical Engineering, Beijing Institute of Petrochemical Technology, Beijing 102617, China

[e]University of Chinese Academy of Sciences, Beijing 100490, China

[*]Corresponding author. Tel/Tax:86-10-82554512. E-mail: bgao@mail.ipc.ac.cn.

[&]These authors contributed to the work equally and should be regarded as co-first authors.

## Abstract

Single pressure refractive index gas thermometry (SPRIGT) is a novel primary thermometry, jointly developed by TIPC of CAS in China and LNE-Cnam in France. To realize a competitive uncertainty of 0.25 mK for thermodynamic temperature measurements, high-stability and low-uncertainty of microwave resonance frequency measurements better than 2 ppb should be achieved. This article describes how to realize high-stability and low-uncertainty of resonance frequency measurements in a copper microwave cavity by experimental optimization methods based on Allan analysis of variance. In this manner, $10^{-12}$ level accuracy and stability of microwave resonance frequency measurements were realized with an integration time of 3 hours, which is nearly 20 times better than those without optimization in our previous work (*Sci. Bull 2019; 64: 286-288*). It has potential applications in gas metrology and other research fields, where high-stability and low-uncertainty microwave measurements are necessary. Besides, microwave measurements were carried out isobarically at pressures of (30, 60, 90, and 120) kPa over the temperature range of (5 to 26) K, with good microwave mode consistency for the determined thermodynamic temperatures. These will provide strong support for the success of the implementation of SPRIGT in China.

2

**1. Introduction**

Single pressure refractive index gas thermometry (SPRIGT), jointly developed by Technical Institute of Physics and Chemistry of Chinese Academy of Sciences (TIPC-CAS) in China and Laboratoire national de métrologie et d'essais-Conservatoire national des arts et métiers (LNE-Cnam) in France [1], is one kind of promising polarizing gas thermometry methods. As a relative primary thermometry [2], SPRIGT measurements are conducted on single isobars rather than isotherms, it decreases the dependence on the accurate absolute pressure measurement and increases the measurement speed ten-fold. With state-of-the-art *ab initio* calculation of helium-4 properties, a competitive uncertainty of 0.25 mK[1] is expected for the measurement of thermodynamic temperature below the neon triple point at 24.5561 K, which is of important strategic significance for the development of both cutting-edge scientific researches and large scientific facilities, such as the Spallation Neutron Source in China and Large Hadron Collider in Europe[3].

Microwave measurements are widely used in the field of gas metrology. In SPRIGT, it allows the thermodynamic temperature to be determined from the ratio of gas refractive indexes at a single pressure, one measured at an unknown temperature and the other at a reference temperature [1, 4], while the refractive index can be measured from the microwave resonance frequencies in a copper quasi-spherical resonator. For the reference temperature, it can be the fixed point of neon or a known thermodynamic temperature measured from other absolute primary thermometry, such as acoustic gas thermometry. In refractive index gas thermometry (RIGT) [2, 5], microwave measurements are also used to get gas refractive index. Then the thermodynamic temperature is deduced from the refractive index and pressure measurements. This has been successfully implemented in NRC between 24.5 K and 84 K at temperatures corresponding to the three ITS-90 defining fixed points, namely, neon, oxygen, and argon [6]. Recent progress on the measurement of the thermodynamic temperature of the triple point of xenon was also realized by further extrapolating the experimentally-determined compressibility at the triple point of water to that at the triple point of xenon [7]. As the reverse application of RIGT, a quantum standard for absolute pressure measurements was jointly developed by LNE in France and INRiM in Italy in the range from 0.2 kPa to 20 kPa by a superconducting microwave cavity [8], where the pressure is deduced from the refractive index and temperature measurements. In acoustic gas thermometry (AGT) [9-12], microwave measurements are used to determine the dimensions or volume of the resonator, the thermodynamic temperature can be then determined by combing the acoustic measurements in the resonator. Besides, microwave measurements are also used in other precise measurements, such as density and critical phenomena of helium [13], frequency [14] and quantum-gas and gravitational physics [15]. These were successfully implemented by using high-quality factor niobium microwave cavities. The above gas metrology (SPRIGT, RIGT, AGT and absolute pressure in a quantum standard) and precise measurements can benefit from the improvement of microwave measurements in reducing the uncertainty component from microwave frequency measurements or improving the long-term stability of the measurements.

---

[1]On May 20th, 2019, the Bureau International des Poids et Mesures announced a major revision to the SI, in which the base units, the kelvin, symbol K, was redefined by fixing the value of Boltzmann constant. The practical realizations of the kelvin by primary thermometry are indicated in the "Mise en pratique for the definition of the kelvin in the SI", in which low-uncertainty primary thermometry is required to promote the realizations of the new kelvin and the spread of high-accuracy, low-temperature metrology. (*see* https://www.bipm.org/utils/en/pdf/si-mep/SI-App2-kelvin.pdf).

Pound-locked loop is typically used in high precision frequency standards [16]. Today the best primary standards can produce the SI second with a relative standard uncertainty almost approaching $10^{-16}$, and the relative uncertainty of secondary frequency standards can be intrinsically accurate at the level of $10^{-18}$ [17]. A convenient way to perform experiments is to use a servo loop to lock the oscillator frequency to the maximum of a resonance. Then the frequency stability depends on the shape and sharpness of the resonance, the signal-to-noise ratio, and the type of noise that predominates at the time scale under consideration [1, 18, 19]. For SPRIGT, in reality, it is essential to perform regular scans of the resonance line to check for any change in the width and shape (see Section 3.1), which will take typically two minutes. To implement SPRIGT, a system has been built at TIPC-CAS in China, which includes three subsystems, namely temperature control system, pressure control system and microwave system. The temperature and pressure control systems are designed to support a very stable working condition (temperature and pressure) for the microwave system. High-stability measurements on temperature control (0.2 mK), pressure control (4 ppm, 1 ppm $\equiv 10^{-6}$) and microwave resonance frequency (2 ppb, 1 ppb $\equiv 10^{-9}$) are required to realize thermodynamic temperature measurements with an uncertainty of 0.25 mK. A simple description of the three subsystems is presented in Section 2.

In this paper, the structure of the quasi-spherical resonator is described and the experimental optimizations on microwave measurement are studied in detail by using Allan analysis of variance, namely, the optimization of microwave emission power, temperature control method and time reference. Besides, frequency measurements were also investigated under four isobars of (30, 60, 90, 120) kPa from 5 K to 26 K. By implementing these microwave optimizations, microwave resonance frequency measurements with accuracy and stability at the $10^{-12}$ level were realized using a -10 dB$_m$ microwave emission power for an integration time of 3 h in the present work. The frequency accuracy and stability are nearly 20 times better than those in our previous work [4]. The result is inspiring, and the optimization methods can be used as a reference to guide how to realize high-stability and low-uncertainty of resonance frequency measurements in a copper microwave cavity. This will improve the measurement accuracy or stability of thermodynamic temperature in SPRIGT, as well as other research fields involving microwave technology, where high-stability and low-uncertainty microwave measurements are necessary.

The rest of the paper is structured as follows. In Section 2, we describe the experimental apparatus involving the main instrumentations for microwave, temperature and pressure measurements. In the present work, gas pressures and the resonator temperatures have been well controlled. A detailed description of the temperature and pressure measurement lies beyond the scope of the present paper, these descriptions can be found in our previous papers [20-22], while the involved improvement on the temperature and pressure systems will be reported elsewhere. In Section 3, we present how to realize high-stability and low-uncertainty of microwave resonance frequency measurement by optimizing the microwave background, the microwave emission power, mechanical stability, temperature control method, time reference, and microwave signal.

## 2. Experimental setup

The apparatus of SPRIGT is shown in Fig. 1a, including microwave, temperature control and pressure control subsystems. For the microwave system, the core element is a copper (Cu-EPT) microwave cavity, *i.e.*, a quasi-spherical resonator (QSR), shown in Fig. 1b. The high-quality factor[2] QSR was built from two hemispheres whose inner surfaces were machined by precision diamond turning. The dimensions were the same as those of the resonator used at LNE-cnam for a determination of the Boltzmann constant [10]. The inner shape is designed to be a triaxial ellipsoid defined by

$$\frac{x^2}{a^2(1+\varepsilon_2)}+\frac{y^2}{a^2}+\frac{z^2}{a^2(1+\varepsilon_1)}=1 \quad (1)$$

with $a$ = 49.50 mm, $\varepsilon_1$ = 0.001, and $\varepsilon_2$ = 0.0005, corresponding to the nominal semi-major axes of the tri-axial ellipsoid of 49.50 mm, 49.75 mm and 50.00 mm on $Y$, $X$ and $Z$ axes, respectively. The nominal shell thickness of the resonator is 10.0 mm. Two loop antennas are connected to a two-port vector network analyzer (Keysight Technologies N5241A PNA-X), one in each hemisphere used for emission and reception. The frequency reference of the vector network analyzer is a 10 MHz signal provided by a rubidium frequency standard (Stanford Research Systems FS725) or GPS time and frequency system (Stanford Research Systems FS740, locked to GPS with OCXO or rubidium frequency standard). For FS740, two kinds of antenna, indoor antenna and outdoor antenna, were used in this work.

The QSR has been closed successfully at room temperature using a microwave method by monitoring the change of relative excess half-width. The resonator is destined to be used under vacuum and then with helium gas. For this reason, to avoid contamination, Argon (purity of 99.999 %, supplied by Beijing AP BAIF Gases

---

[2]Quality factors ($Q = f/2g$) for the 4 microwave modes at 5 K ~26 K are: $Q$(TM11)≈140000, $Q$(TE11)≈210000, $Q$(TM12)≈170000, $Q$(TE13)≈290000.

Industry Co. Ltd.) was flowed through it at a continuous rate of 80 SCCM³ until the resonator was coupled in the pressure vessel of the cryostat. After putting the resonator into the pressure vessel of the cryostat, a preliminary test on the performance of the quasi-spherical resonator under vacuum was implemented without using any optimization methods involved in the present work. Frequency stability and uncertainty of 0.02 ppb and 0.03 ppb were realized by using 0 dB$_m$ microwave emission power with an integration time of 3 h [4], equivalently 0.063 ppb and 0.095 ppb for -10 dB$_m$ microwave emission power [1].

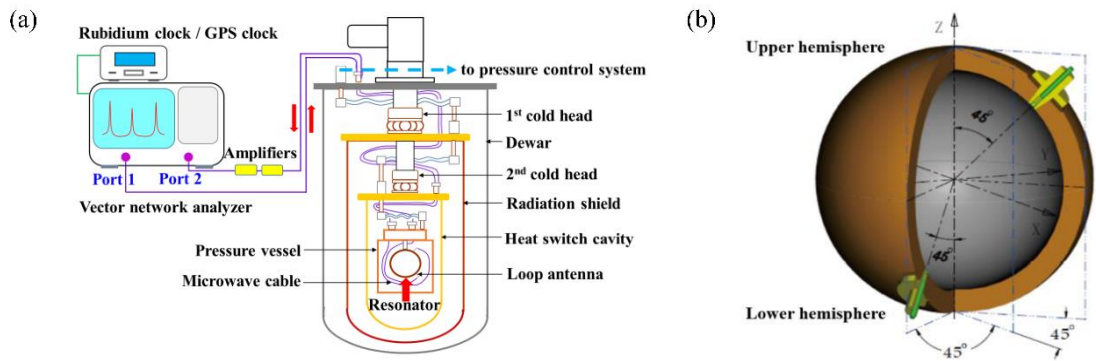

Fig.1. The structure schematic diagram of SPRIGT. (a) Simplified system diagram. (b) 3D drawing of the quasi-spherical resonator.

For the temperature control system, a cryogen-free cryostat was developed for SPRIGT in our previous work [20], using a two-stage GM type pulse tube cryocooler as the cooling resource. Based on multi-layer radiation shields combined with the thermal-resistance method, gas type heat switch and PID control method, the pressure vessel temperature stabilities of 0.021 mK ~ 0.050 mK were realized with an integration time of 0.8 s in the temperature range from 5 K to 25 K. Based on further investigation on thermal characterization of the cryostat, the pressure vessel temperature stabilities were controlled at 0.019 mK in the same range [21]. In this work, the temperature of the resonator was measured by a rhodium iron thermometer (Tinsley, SN226242) coupled to an AC resistance bridge (ASL F900). The 10 Ω standard reference resistance (Tinsley 5685A, SN1580409) was placed in an oil bath (Aikom Instruments MR 5100-L) with temperature stability better than 1 mK.

---

³SCCM = standard cubic centimeters per minute, corresponding to 1.6667×10⁻⁸ m³·s⁻¹ in the International System of Units (SI). We define the volumetric flow 1 sccm as the flow of 1 cubic centimeter per minute of argon at the pressure 101.325 kPa and temperature 20℃.

For the pressure control system, the detailed content is not emphasized in this work since it has been presented in our previous work [22]. The uncertainty contribution of pressure to measurements of thermodynamic temperature has also been reported in our previous work [1] and Fig. 2 in the review paper of RIGT [2]. To fill helium-4 gas (purity of 99.9999 %, supplied by Air Liquide) into the resonator and keep a single pressure, a gas compensation loop was built at room temperature to compensate for the leak of the piston gauge. The gas pressure was measured by an absolute-pressure piston gauge (Fluke PG 7601) and conducted by maintaining the piston at a constant height using a laser interferometer (KEYENCE LK-G80). Pressure stabilities of 0.0032 Pa at 30 kPa and 0.002 Pa at 90 kPa were realized with an integration time of 1.4 s at room temperature, corresponding to relative stabilities of 0.1 ppm and 0.02 ppm, respectively [22].

## 3. Experimental optimization and discussions

To realize high-stability and low-uncertainty microwave resonance frequency measurements at low temperatures, several experimental optimizations were implemented by using Allan analysis of variance in the present work. Detailed results are listed as below.

*3.1. Microwave background polynomial*

In this work, the microwave resonance frequencies $f_n$ and half-widths $g_n$ were determined from non-linear least-squares fitting (LM method [23]) of the measured complex scattering parameters $S_{21}$ as a function of frequency [24]. Here we rewrite equation 15 in the literature to the following form

$$S_{21} = \sum_n \frac{A_n f}{f^2 - (f_n + ig_n)^2} + \sum_{j=0}^{J} B_j (f - f_*)^j \qquad (2)$$

where the fitting parameters were the complex constants $A_n$, $B_j$ and the three complex resonance frequencies $(f_n + ig_n)$, one for each component of the triplet. In equation 2, $J$ is the background polynomial order, $f$ is the source frequency and $f_*$ is an arbitrary constant, here we made the same choice as the literature [24], *i.e.*, $f_* = f_x$, to avoid numerical problems in the fitting program.

Fig.2 shows the fitting results of TM11 microwave mode for different background polynomial orders at 5 K and 24.5 K under vacuum. With the polynomial order increasing, the average fitted triplet frequency $\langle f + g \rangle$ and the uncertainty $u(\langle f + g \rangle)$ have a little change, as shown in Fig. 2a and Fig. 2c; the maximum relative changes on $\langle f + g \rangle$ are about $2.0 \times 10^{-10}$ and $1.9 \times 10^{-10}$ for 5 K and 24.5 K, respectively. These changes are less than half of the relative standard uncertainty of $\langle f + g \rangle$, indicating that increasing the microwave background polynomial order cannot obviously reduce their influence on the average fitted triplet frequency and the uncertainty in our system. Considering other microwave modes and those measured at room temperatures, a 2$^{nd}$-order background polynomial is competent in this work. Table 1 lists the fitted frequencies and half-widths of the triplet for 2$^{nd}$-order background polynomial. Fig. 2b and Fig. 2d compare the results of the measured and calculated components of the scattering parameters $S_{21}$, where good agreement was found.

Table 1. Fitted values of the triplet with 2$^{nd}$-order background polynomial.

| $T$ / K | $f_1$ / MHz | $g_1$ / MHz | $f_2$ / MHz | $g_2$ / MHz | $f_3$ / MHz | $g_3$ / MHz |
| --- | --- | --- | --- | --- | --- | --- |
| 5 | 2617.0688336(11) | 0.0093815(11) | 2617.6300911(15) | 0.0093725(15) | 2618.2397203(16) | 0.0098880(16) |
| 24.5 | 2617.0614026(12) | 0.0094027(12) | 2617.6226012(16) | 0.0093916(16) | 2618.2321434(17) | 0.0099124(17) |

Besides, the background noise from cables and feed-throughs also influences the microwave measurement. To reduce the noise as much as possible, one can use some low-loss cables and feed-throughs, covering the working frequency range. For further improvement, the optimal combination of cables and feed-throughs, even the working microwave modes can be selected by measuring and analyzing the $S_{21}$ parameters.

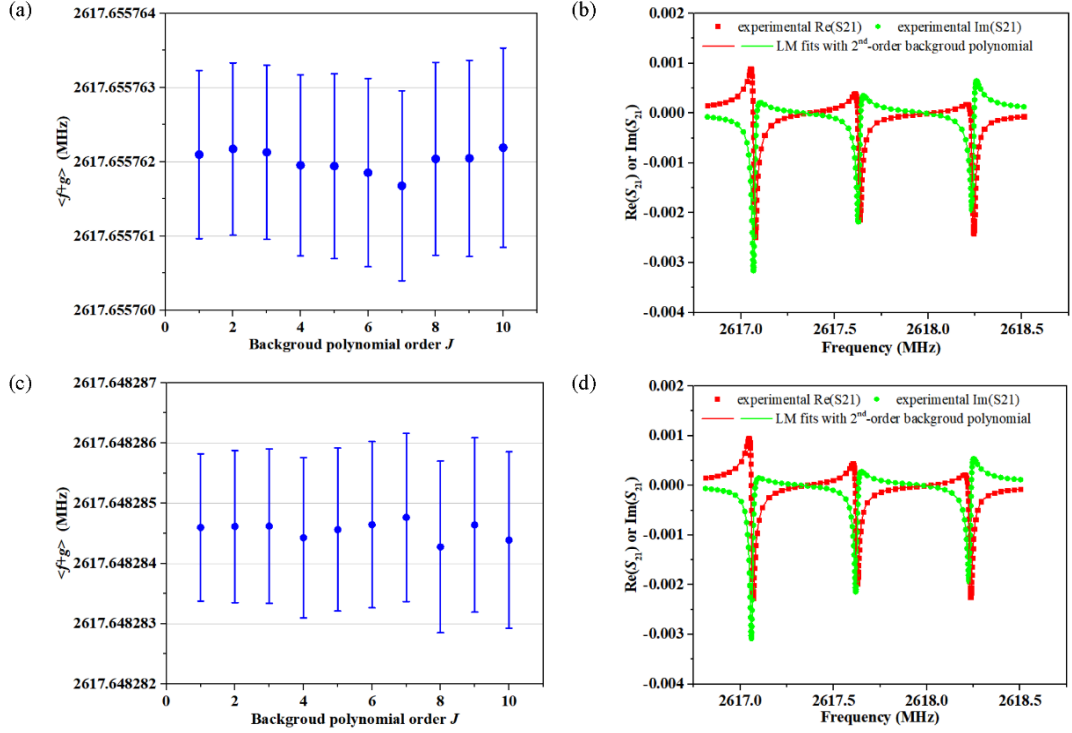

Fig.2. Fitting results of TM11 microwave mode for different background polynomial orders under vacuum: (a) $<f + g>$ against microwave background polynomial order $J$ at 5 K. (b) Real and imaginary components of the parameter $S_{21}$ at 5 K, measured and fitted using $2^{nd}$-order background polynomial. (c) $<f + g>$ against microwave background polynomial order $J$ at 24.5 K. (d) Real and imaginary components of the parameter $S_{21}$ at 24.5 K, measured and fitted values using $2^{nd}$-order background polynomial.

*3.2. Microwave emission power*

In this work, the microwave signal was generated by a two-port N5241A vector network analyzer. The signal was transmitted from port 1 of the vector network analyzer, passing through the cables, feed-throughs, and resonator, later received by port 2. Since the relative loss of the cables and feed-throughs are nearly constant at stable operating conditions, the microwave emission power has a direct influence on the measurement inside the resonator. If the power is too small, the microwave signal may be drowned in the background noise and thus leads to a relatively larger uncertainty of the fitted frequency or even no triplet. On the contrary, the microwave signal will heat the resonator too much, this will increase the temperature control instability and in turn reduce the frequency stability. Thus, it is necessary to optimize the microwave emission power.

Fig. 3 shows the optimized results of the power at 5.0 K under vacuum based on the TM11, TE11, TM12 and TE13 microwave modes, which were employed in our preliminary measurements [4]. A bigger power is easier to increase the temperature instability, as shown in Fig. 3a and Fig. 3b, where the temperature stabilities are about 0.042 mK and 0.024 mK for microwave emission power $P_{MW} = -7$ dB$_m$ and $P_{MW} = -10$ dB$_m$, respectively. Fig. 3c plots the temperature Allan standard deviation against the integration time for these two powers. It can be seen that there is an oscillation of the temperature Allan standard deviation for $P_{MW} = -7$ dB$_m$, which is coupling with the measurements of the 4 microwave modes, measured in the TM11-TE11-TM12-TE13 loop. While for $P_{MW} = -10$ dB$_m$, no oscillation was observed on the temperature Allan standard deviation, which means the microwave power has negligible disturbance effect on temperature stability. Thus the optimal microwave emission power was found to be $P_{opt} = -10$ dB$_m$. In the present work, we use one single microwave emission power $P_{opt}$ during all the low-temperature microwave frequency measurements no matter whether the resonator is under vacuum or filled with high-purity helium-4 gas.

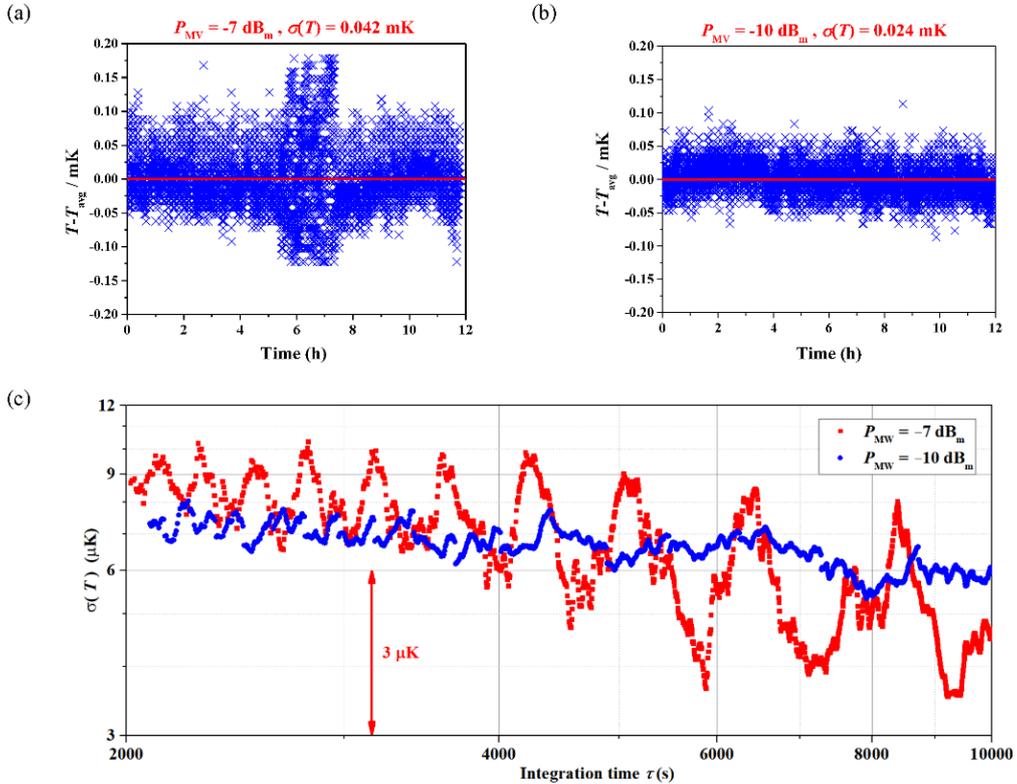

Fig.3. Microwave emission power optimization at 5.0 K under vacuum. (a) Temperature evolution with $P_{MW} = -7$ dB$_m$. (b) Temperature evolution with $P_{MW} = -10$ dB$_m$. (c) Temperature Allan standard deviation as a function of integration time for $P_{MW} = -7$ dB$_m$ and $P_{MW} = -10$ dB$_m$.

*3.3. Mechanical stability, temperature control method, time reference, and signal intensity*

In this section, mechanical stability, temperature control method, time reference and signal intensity were also studied in this work under vacuum unless otherwise specified.

*3.3.1. Mechanical stability*

When cooling the resonator from room temperature down to low temperatures (5 K to 25 K) for each independent run in this work, it is important to keep the resonator in good mechanical stability before doing the microwave measurements. A simple way to release the mechanical stress is to do temperature cycling in the objective temperature range, for example, cooling and warming between 25 K and 5 K in this work. Because releasing mechanical stress needs some time. Besides, it is also good for temperature and frequency measurements. At the same set point for temperature control, after the temperature cycling, it is easier to realize highly stable temperature measurements, which has been observed in our system. The temperature stability, in turn, will improve the frequency measurement as shown in Section 3.3.2 below. When the temperature of the resonator reached the objective temperatures (5 K and 24.5 K), temperature regulations together with microwave measurements were implemented for several hours. If the frequencies of each microwave mode had no obvious change anymore (changing within microwave measurement uncertainty) at the objective temperatures, then it can be considered that the mechanical stress of the resonator is completely released. A cooling-warming temperature cycling for 55 hours was carried out in this work. The result is shown in Fig. 4a, where the cooling process and warming process are denoted as C and W, respectively, in the figure legend, and the behind number is the sequence of the temperature cycle. The average amplitude of $|A_p|$ in equation 2 of the cooling process and warming process, denoted as $A$, was used to indicate the frequency change with the temperature change. They do not completely overlap mainly because of a big difference in temperature change speed between them. This will be improved later in our work by controlling the cooling and warming speed automatically. Generally, after several cooling and warming cycles, microwave measurements can be implemented in the resonator when the temperature and the filled gas pressure are stable simultaneously. This is the case for the present work, where obvious resonance frequency changes were not observed for each microwave mode at the two objective temperatures, respectively.

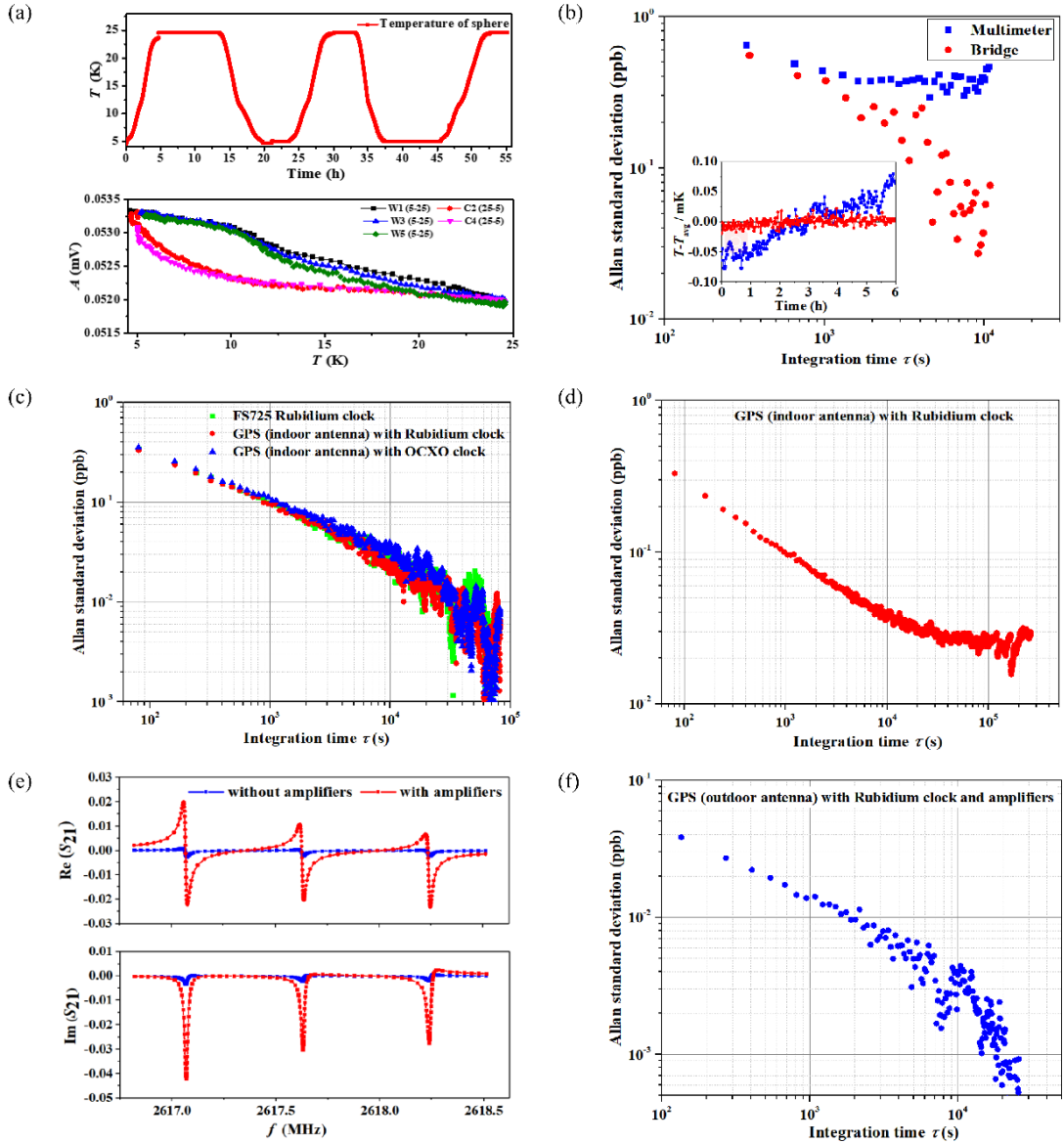

Fig.4. Optimized results of mechanical stability, temperature control, time reference and signal on TM11 microwave mode under vacuum unless otherwise specified. (a) Evolution of resonator temperature and average amplitude of $|A_p|$ in equation 2 during temperature cycle. (b) Frequency

Allan standard deviation with two temperature control methods at $T$ = 15.0 K and $p$ = 30 kPa. (c) Two-days frequency stability at 24.5 K with different time references. (d) Seven-days frequency stability at 5.0 K for FS740 time reference locked to GPS with Rubidium atomic clock configured indoor antenna. (e) Real and imaginary components of $|S_{21}|$ of the triplet at 5.0 K with and without the two microwave amplifiers. (f) Frequency Allan standard deviation at 5.0 K using FS740 locked to GPS with Rubidium atomic clock configured outdoor antenna and amplifiers.

*3.3.2. Temperature control method*

Due to thermal expansion effect of the resonator, its dimensions will change if the temperature is not stable. The microwave resonance frequency is proportional to the reciprocal of the radius of the resonator, thus the temperature stability has a direct influence on microwave measurements. Fig. 4b shows the influence on microwave resonance frequency stability of TM11 mode for two temperature control methods at $T$ = 15.0 K and $p$ = 30 kPa. Method 1 uses multimeter (Keithley 2002) with Cernox sensor as in our previous work [20], while method 2 uses AC resistance bridge (ASL F18) with rhodium iron sensor as in CCT-K1 [25]. The temperature stabilities of the two methods are 38 μK and 7 μK with an integration time of 33.6 s, respectively. The frequency stability of method 2 is better than that of method 1 by about an order of magnitude with an integration time $\tau$ = 3 h. Good temperature stability can improve the microwave resonance frequency stability. Thus, in our later measurements, method 2 was adopted all the time to maintain a good frequency stability.

*3.3.3. Time reference*

Usually, the vector network analyzer is connected to an external time reference, such as rubidium atomic clock or GPS clock, to realize high-accuracy and high-stability frequency measurements. In this work, rubidium atomic clock (FS725) and GPS clock (FS740 locked to GPS) were used. While the long-term stability of GPS is excellent, its short-term stability is rather poor in comparison to modern oscillators [26]. So in this work, two different oscillators, an OCXO and a rubidium atomic frequency standard, were configured for FS740. Fig. 4c shows the typically measured stability of microwave resonance frequency of TM11 mode at 24.5 K in two days with the three kinds of time references, namely, FS725 (rubidium atomic clock), FS740 locked to GPS with rubidium atomic clock and with OCXO clock, with GPS configured indoor antenna. The frequency stability of FS740 locked to GPS with rubidium atomic clock configured indoor antenna and that of the FS725 are comparatively equivalent, both stabilities are better than that of OCXO clock. A resonance frequency stability of 0.02 ppb was realized using FS740 locked to GPS with rubidium atomic clock configured indoor antenna and FS725 with an integration time $\tau$ = 3 h in the two-days microwave measurements. The SPRIGT measurement at different temperatures and pressures may take one or two months, the time reference stability of the reference should be tested at this level to check if any other effect may perturb the measurement. Furthermore, long-term microwave measurements at 5 K in seven days were carried out by FS740 locked to GPS with rubidium atomic clock configured indoor antenna. The stability result is plotted in Fig. 4d. A good resonance frequency stability of 0.04 ppb was achieved with $\tau$ = 3 h.

*3.3.4. Signal intensity*

Amplifier is usually used to enhance the microwave signal intensity as shown in literature [27], where the amplitude of $|S_{21}|$ was amplified by nearly ten times with the help of a set of two 10 dB$_m$ amplifiers (Model ZX60-14012L manufactured by Mini-Circuits, bandwidth from 0.3 MHz to 14 GHz) connected in series at room temperature. The uncertainty was reduced by a factor of 5 - 12 even with a low-cost vector network analyzer. In this work, with the same method and two amplifiers like the literature [27], measurements were carried out at 5 K under vacuum. Fig. 4e shows the comparison results of the real and imaginary components of $|S_{21}|$ with and without the two amplifiers. With amplifiers, the amplitudes were amplified by about 10 times and in turn, the fitted relative standard uncertainty of resonance frequency was reduced by a factor of 10 (from 0.3 ppb to 0.031 ppb). This is because the amplifiers increase the resolution of the peaks of the triple. A similar situation appears in TE11, TM12, and TE13 microwave modes, with the fitted relative standard uncertainty of resonance frequency reduced by a factor of 5, 5, and 10, respectively. It is a simple and effective way to solve signal problems, such as high-loss of microwave cable and feed-through, even low-cost vector network analyzer. Besides, low-temperature microwave amplifiers are expected to have a better performance than room-temperature amplifiers, and this will be one solution of ultra-high accuracy microwave measurements in the future work.

In Section 3.3.3, the microwave measurements were carried out by using the indoor antenna, which was placed outside a window of the laboratory building, facing north against another building. Actually, the FS740 was supplied with two types of antenna, indoor antenna and outdoor antenna. According to the user's manual of the FS740, best results can be realized if the antenna has a clear unobstructed view of the sky. It is highly recommended to put the outdoor antenna on the roof of the building, within which the FS740 is located. Because it aggrandizes the quality and reliability of the GPS signals as more satellites will typically be visible and with more signal noise ratio. This can aggrandize the long-term stability of the FS740 by a factor of three. Thus, in the following test of this Section, the outdoor antenna was employed and placed on the roof of the laboratory building. Combining the above optimized results, namely, 2$^{nd}$-order background polynomial, optimized power $P_{opt} = -10$ dB$_m$, good mechanical stability, temperature control with AC resistance bridge and rhodium iron sensor, time reference FS740 locked to GPS with rubidium atomic clock but configured outdoor antenna and two room-temperature amplifiers connected in series, preliminary stability measurements on TM11 mode were implemented at 5 K under vacuum. Fig. 4f plots the Allan standard deviation of the resonance frequency. With the same time reference, FS740 locked to GPS with rubidium atomic clock, the stability for outdoor antenna with amplifiers is about an order of magnitude better than that for only indoor antenna as shown in Fig. 4d and Fig. 4f. High-stability of 3.7 parts per trillion (ppt, 1 ppt ≡ $10^{-12}$ ≡ $10^{-3}$ ppb) and low-uncertainty of 5.2 ppt were realized with an integration time of 3 h. The frequency stability and uncertainty in the present work are nearly 20 times better than those in our previous work [4], 0.064 ppb and 0.095 ppb with the same integration time and $P_{opt} = -10$ dB$_m$ [1]. In SPRIGT, the thermodynamic temperature uncertainty has many influence factors, frequency is one of the major factors as shown in our previous work [1] and the RIGT review paper [2]. The improvement in frequency measurement accuracy can reduce its influence on thermodynamic temperature measurement by a factor of 20, which is good for realizing the high-accuracy measurement of thermodynamic temperature by SPRIGT. The above optimization methods have the potential to be used in gas metrology and other research fields, where high-stability and low-uncertainty microwave measurements are necessary.

*3.4. Results with pressures*

After many experimental optimizations under vacuum, preliminary microwave measurements were performed from 5 K to 26 K with the resonator filled by high-purity helium-4 at the following conditions: 1) fitting $S_{21}$ with 2$^{nd}$ order background polynomial; 2) setting microwave emission power $P = P_{opt} = −10$ dB$_m$; 3) doing temperature cycle to make the resonator with a good mechanical stability; 4) controlling temperature by AC resistance bridge and rhodium iron sensor; 5) using time reference FS725; and 6) without the two amplifiers. The measurements were implemented on isobars and the results are described in the Fig. 5.

Fig. 5a shows the total relative standard uncertainty of <*f*+*g*> of TM11 mode at temperatures from 5 K to 26 K and pressures up to 120 kPa with $\tau = 80$ s[4]. Most uncertainties are located within 0.9 ppb, except the *p-T* region with pressures near 120 kPa and temperatures near 5 K. The main reason is that at lower temperatures, the pressure stability inside the resonator becomes a little worse than that at higher temperatures, which disturbed the frequency stability by the change of helium-4 density inside the resonator. Fig. 5b plots a one-day microwave measurement of TM11 mode at 24.5 K with different pressures. The frequency stabilities are about (0.03, 0.02, 0.04, 0.05) ppb for (30, 60, 90, 120) kPa with $\tau = 3$ h, respectively, and the temperature stabilities are about (6.9, 8.3, 7.9, 6.6) μK with $\tau = 33.6$ s. Due to the pressure stability, standard uncertainty of <*f* + *g*> for TM11 mode with $\tau$ = 80 s. (b) One-day frequency stability measurements at 24.5 K for (30, 60, 90, 120) kPa. (c) Thermodynamic temperature differences ($T_{\text{mode}} - T_{\text{avg}}$) for TM11, TE11 and TE13 microwave modes at 60 kPa. (d) Standard uncertainty component of SPRIGT thermodynamic temperature from mode consistency.

---

[4]It is usually used a constant integration time for all the microwave modes. Figure 5(a) only shows the uncertainty of TM11 microwave mode for different pressures and temperatures. However, for TE13 mode, the uncertainties are very near 2 ppb. Besides, the integration time was also chosen to be in the same "time" as the other instruments, such as the resistance bridge. The integration time of 20000 s is used to check and characterize the stability of the measurements.

60 kPa has the best frequency stability in the one-day stability measurement. These are still better than our previous work (0.063 ppb under vacuum provided with $P_{opt}$ = −10 dBm) even the pressure instability increases the microwave frequency measurement instability. The stability and uncertainty still have room for improvement, provided later using the time reference, FS740 locked to GPS with rubidium atomic clock configured outdoor antenna, and microwave amplifiers. Besides, thermodynamic temperatures calculated from different modes were determined for the isobar $p$ = 60 kPa, based on the SPRIGT principle [1]. Fig. 5c and Fig. 5d presents the differences between mode thermodynamic temperatures $T_{mode}$ and the average values $T_{avg}$, ($T_{mode}$ − $T_{avg}$), at temperatures from 5 K to 26 K, where microwave measurements with $\tau$ = 80 s were used. Good mode consistency was observed on ($T_{mode}$ − $T_{avg}$) within ±80 μK, and a small relative standard uncertainty component for the thermodynamic temperatures, from mode consistency, was realized with $u(T_{mode})$ values less than 60 μK[5].

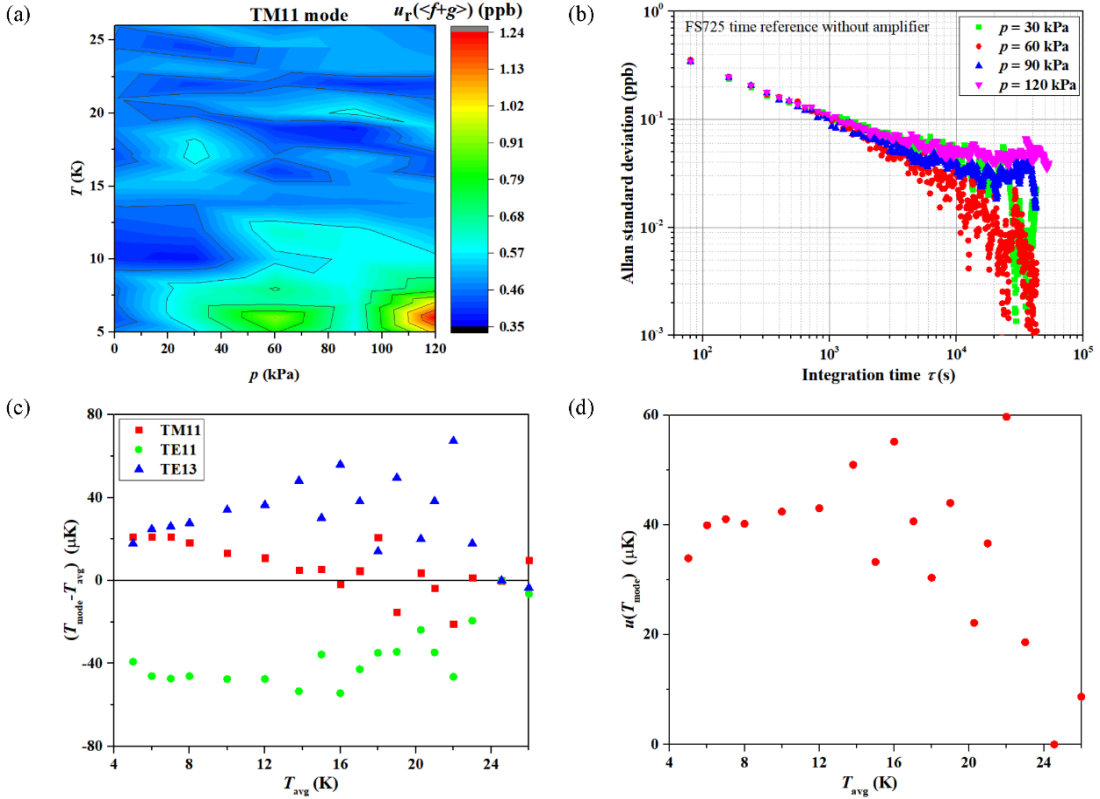

Fig.5. Microwave measurements from 5 K to 26 K at pressures up to 120 kPa. (a) Total relative

---

[5]Uncertainty budget for SPRIGT has been reported in our previous work [1] and the review paper of RIGT [2]. Since this work is focused on microwave measurements, it will be conducted in our future work.

## 4. Conclusion

In this study, experimental optimizations on microwave measurements were demonstrated at low temperatures. By doing this, the best available combination for our microwave system was determined: 1) fitting $S_{21}$ parameters with 2$^{\text{nd}}$ order background polynomial; 2) setting microwave emission power $P = P_{\text{opt}} = -10$ dB$_{\text{m}}$; 3) doing temperature cycle to make the resonator with a good mechanical stability; 4) controlling temperature by AC resistance bridge and rhodium iron sensor; 5) using time reference FS740 locked to GPS with rubidium atomic clock configured outdoor antenna; 6) with room-temperature microwave amplifiers, improvements are still expected with the implementation of low-temperature amplifiers later. In this manner, microwave resonance frequency measurements with accuracy and stability at the $10^{-12}$ level were realized at low temperatures in our quasi-spherical resonator. The frequency stability and accuracy in this work were enhanced by nearly 20 times compared with those in our previous work without optimization. The present optimization methods can be used as a reference to guide how to realize high-stability and low-uncertainty frequency measurements in a copper microwave cavity, not only for improving thermodynamic temperature measurement in SPRIGT, but also for other precise measurements, where high-stability and low-uncertainty microwave measurements are necessary. Besides, the thermodynamic temperatures were calculated for the isobar 60 kPa, based on the principle of single pressure refractive index gas thermometry. Good mode consistency and lower uncertainty were achieved. The present work should provide strong support for the successful establishment of single pressure refractive index gas thermometry in China.

**Acknowledgments**

This work is supported financially by the National Key R&D Program of China (Grant No. 2016YFE0204200), the National Natural Science Foundation of China (Grant No. 51627809), the International Partnership Program of the Chinese Academy of Sciences (Grant No. 1A1111KYSB20160017), and the EMRP project Real-K (No. 18SIB02). Author Changzhao Pan was supported by the funding provided by the Marie Skłodowska-Curie Individual Fellowships-2018 (834024). The authors gratefully acknowledge Richard Rusby from National Physical Laboratory UK for sharing his long experience and constant advice on temperature measurements. We are deeply grateful to Wei Wu from City University of Hong Kong China for his kindly reading and helpful suggestions on the manuscript. We would also like to thank Kun Liang from the National Institute of Metrology China and Chongxia Zhong from Beijing Institute of Metrology China for their helpful discussion on time reference.